\documentclass{ws-ijmpd}

\begin{document}

\markboth{Roland Triay}{Roland Triay}

%
\catchline{}{}{}{}{}
%

\title{AN ISSUE TO THE COSMOLOGICAL CONSTANT PROBLEM}

\author{ROLAND TRIAY}

\address{Universit\'e de Provence\\ and Centre de Physique Th\'eorique CNRS (UMR\,6207)\\ Luminy Case
907, F~13288 Marseille Cedex 9, France\\
triay@cpt.univ-mrs.fr}

\maketitle

\begin{history}
\received{Day Month Year}
\revised{Day Month Year}
\comby{Managing Editor}
\end{history}

\begin{abstract}
According to general relativity, the present analysis shows on geometrical grounds that the cosmological constant problem is an artifact
due to the unfounded link of this fundamental constant to vacuum energy density of quantum fluctuations.
\end{abstract}


\section{Introduction}	
The status of the cosmological constant $\Lambda$ has long been
discussed\cite{Souriau77,FeltenIsaacman86,CharltonTurner87,Sandage88}, whereas it is clearly established in General
Relativity (GR) as {\em universal constant\/}\cite{Souriau64}. Therefore, similarly to Newton constant of gravitation
${\rm G}$, its value has to be estimated from observations.  However, such an estimate does not agree by hundred orders
of magnitude with its expected value as obtained from quantum field theories\cite{Carroll01,Straumann02,Padmanabhan03}
by assuming that vacuum energy density of quantum fluctuations is the origin of this constant. The aim of the present
analysis is to analyse on geometrical grounds this problem, called {\em cosmological constant problem\/} (CCP).

\section{Status of the cosmological constant}
The cosmological constant was assumed in the field equations for describing the observations in accordance with a static
cosmological solution\cite{Einstein17} but a general expansion of the universe was observed\cite{Hubble27} subsequently.
What is usually called ``Einstein's biggest blunder'' stands probably for the historical reason why $\Lambda$ was wrongly
understood as a free parameter in the field equations (see\cite{Straumann02,Padmanabhan03} for more details). Such an
issue to the cosmological problem has provided us with (authority and/or simplicity) arguments\cite{Einstein31} in favor
of $\Lambda=0$ until acceleration of the cosmological expansion could not be avoided for the interpretation of recent
data (chap.\,\ref{Observations}). On geometrical grounds, the {\em principle of general relativity\/}
(PGR) applied to gravity provides us with the status of universal constant for $\Lambda$, which intervenes in the
description of the gravitational field at cosmological scales (chap.\,\ref{GravityGR}), similarly as for ${\rm G}$ at
smaller scales.

\subsection{Observational status of $\Lambda$}\label{Observations}
In the past, estimates such as $\Lambda<2\,10^{-55}\,{\rm cm}^{-2}$ from dynamics of galaxies in clusters\cite{Peach70} or $-2\,10^{-56}\,{\rm
cm}^{-2}\leq\Lambda<4\,10^{-56}\,{\rm cm}^{-2}$ from the minimum age of the universe and the existence of high redshift objects\cite{Carroll92}, were
interpreted with some {\em a priori\/} in mind (for arguing) in favor of a vanishing value. Decades later, estimates based on the redshift--distance
relation for brightest cluster galaxies\cite{BigotEtal88,BigotTriay89} and for quasars\cite{FlicheSouriau79,FST82,BigotEtal88,Triay89} provided
us unambiguously with a non zero cosmological constant $\Lambda\sim 3h^{2}\,10^{-56}\,{\rm cm}^{-2}$, where
$h=H_{\circ}/100\,km\,s^{-1}\,Mpc^{-1}$. Nowadays, it is generally believed that $\Lambda\sim 2h^{2}\,10^{-56}\,{\rm
cm}^{-2}$ is required for interpreting the CMB temperature
fluctuations\cite{SieversEtal02,NetterfieldEtal02,SpergelEtal03,BenoitEtal03} and for accounting of Hubble diagram of
SN\cite{PerlmutterEtal98,PerlmutterEtal99,SchmidtEtal98,RiessEtal98}.

\subsection{Geometrical status of $\Lambda$}\label{GravityGR}
The gravitational field and its sources are characterized respectively by the metric tensor $g_{\mu\nu}$ on the space-time
manifold $V_{4}$ and by a {\em vanishing divergence\/} stress-energy tensor $T_{\mu\nu}$. The gravitational field equations satisfy PGR~: they
must be invariant with respect to the action of diffeomorphism group of $V_{4}$\cite{Souriau64,Souriau74}. In other
words, their most general form reads as an expansion of covariant tensors written in term of the metric tensor
$g_{\mu\nu}$ and it derivatives as follows
\begin{equation}\label{FundEq}
T_{\mu\nu}=-A_{0}F_{\mu\nu}^{(0)}+A_{1}F_{\mu\nu}^{(1)}+A_{2}F_{\mu\nu}^{(2)}+\ldots
\end{equation}
where $F_{\mu\nu}^{(n)}$ are tensors of order $2n$ and $A_{n}$ is a {\em coupling constant\/}. The tensors of order
$\leq 2$ are uniquely defined,
\begin{equation}\label{ET}
F_{\mu\nu}^{(0)}=g_{\mu\nu},\quad
F_{\mu\nu}^{(1)}=S_{\mu\nu}=R_{\mu\nu}-\frac{1}{2}Rg_{\mu\nu}
\end{equation}
where $R_{\mu\nu}$ stands for the Ricci tensor and $R$ the scalar curvature, whereas $F_{\mu\nu}^{n\geq2}$
must be derived from additional principles. The values of coupling constants $A_{n}$ must be estimated from observations.

Schwarzschild solution of Eq.\,(\ref{FundEq}) enables us to identify $A_{n=0,1}$ with Newton approximation, what provides us
with modified Poisson equation\cite{Souriau64}
\begin{equation}\label{Poisson}
\rm{div}\vec{g}=-4\pi {\rm G}\rho + \Lambda
\end{equation}
where $\vec{g}$ stands for the gravitational acceleration field due to sources defined by a specific density $\rho$, and
the following identification of constants
\begin{equation}\label{Const}
{\rm G}=\frac{1}{8\pi A_{1}},\qquad \Lambda = \frac{A_{0}}{A_{1}}
\end{equation}
which shows their common status of {\em universal constant\/}. Therefore, one understands that the same treatment has to
be applied to both of them for estimating their values from observations but at scales adapted to each of them, as it can
be shown from a dimensional analysis of Eq.~(\ref{FundEq},\ref{ET}).

According to GR, the speed of the light $c=1$ ({\it i.e.\/} time can be measured in unit of length\footnote{This is the reason why any statement on
$c$ is meaningless in GR ({\it e.g.\/} to be variable).} $1{\rm s}=2.999\,792\,458\,10^{10}\,{\rm cm}$) and then ${\rm
G}=7.4243\times10^{-29}\;{\rm cm}\,{\rm g}^{-1}$. Let us choose units of mass and of length\footnote{Only two fundamental units can be chosen, the
third one is derived.}, herein denoted respectively by $M$ and $L$. The correct dimensional analysis of GR sets the covariant metric tensor to have
the dimension
$[g_{\mu\nu}]=L^{2}$, and thus
$[g^{\mu\nu}]=L^{-2}$, $[R_{\mu\nu}]=1$ and $[R]=L^{-2}$. Since the specific mass density and the pressure belong to $T^{\mu}_{\nu}$, one has
$[T_{\mu\nu}]=ML^{-1}$. Hence, according to Eq.\,(\ref{FundEq}), the dimensions of $A_{n}$ are the following
\begin{equation}\label{DimCst}
[A_{0}]= ML^{-3},\quad [A_{1}]= ML^{-1},\quad \ldots [A_{n}]= ML^{2n-3}
\end{equation}
which shows their relative contributions for describing the gravitational field with respect to scale. Namely, the larger their order $n$ the
smaller their effective scale\footnote{In other words, the contribution of $A_{0}$ dominates at scale larger than the one of
$A_{1}$, {\it etc\/}\dots}. Equivalently, the estimation of $A_{0}$ demands observational data located at scale larger than the one for
$A_{1}$, {\it etc\/}\dots. This is the reason why the $\Lambda$ effect is not discernible at small scale but requires cosmological distances.

\subsection{Modeling gravitational structures}\label{ModGravStruct}

The space-time geometry is constrained by the presence of gravitational sources as described by means of tensor $T_{\mu\nu}$ in
Eq.~(\ref{FundEq}). According to dimensional analysis given in previous subsection, each right hand terms contributes for describing the geometry
within its effective scale. The observations show that gravitational structures within scales of order of solar system can be described by limiting
the expansion solely to Einstein tensor $S_{\mu\nu}$, when cosmology requires also the first term. The transition scale between $A_{0}$ and $A_{1}$ is
of order of $1/\sqrt{\Lambda}\sim7h^{-1}\;{\rm Gyr}$. Although GR is preferred for investigating the dynamics of cosmic structures, Newton
approximation given in Eq.\,(\ref{Poisson}) provides us with an easier schema for realizing the $\Lambda$ effect. Hence, the acceleration
field due to gravity around a point mass $m$ reads
\begin{equation}\label{g}
\vec{g}=\left(-{\rm G}\frac{m}{r^{3}} + \frac{\Lambda}{3}\right)\vec{r}
\end{equation}
Since $\Lambda>0$, the gravity force is attractive at distance $r<r_{\circ}$ and repulsive at $r>r_{\circ}$ with a critical distance
\begin{equation}\label{r0}
r_{\circ}=\sqrt[3]{3m{\rm G}/\Lambda}
\end{equation}
where the gravity vanishes. In accordance with observations, no $\Lambda$ effect is expected in the sun neighborhood because
$r_{\circ}\sim10^{2}h^{-2/3}\;{\rm yr}$ is much larger than the size of solar system and the mean distance between stars. On the other
hand, it should be appreciable in the outer parts of the Galaxy since $r_{\circ}\sim5\,10^{5}h^{-2/3}\;{\rm yr}$ is only 5
times larger than the disc diameter\footnote{With this in mind, the dynamics of the extended HI regions of spiral galaxies should be
reviewed with respect to the interpretation of rotation curves.}. In the case of Local Super Cluster,
$r_{\circ}\sim4\,10^{8}h^{-2/3}\;{\rm yr}$ corresponds approximately to its size, what suggests that a $\Lambda$ effect might intervenes in its
formation process. The hypothesis that the value of $\Lambda$ accounts for the smoothing scale $\sim$100\,Mpc from which the distribution of
cosmological structures becomes homogeneous and isotropic today should be envisaged. 

\section{The cosmological constant problem}

It is assumed that the contribution of quantum fluctuations to the gravitational field is defined by the following stress-energy tensor\footnote{The
usual picture which describes the vacuum as an isotropic and homogenous distribution of gravitational sources with energy density $\rho_{\rm vac}$ and
pressure $p_{\rm vac}=-\rho_{\rm vac}$ (although this is not an equation of state) is not clear and not necessary for the discussion. }
\begin{equation}\label{TVacuum}
T_{\mu\nu}^{\rm vac}=\rho_{\rm vac}\,g_{\mu\nu},\qquad \rho_{\rm vac}=\hbar k_{\rm max}
\end{equation}
in the field equations\,Eq.(\ref{FundEq}), where $k_{\rm max}$ stands for the ultraviolet momentum cutoff up to which the
quantum field theory is valid\cite{Carroll01}. However, the expected density, {\it e.g.\/}
\begin{equation}\label{DensityQFT}
\rho_{\rm vac}^{EW}\sim 2\,10^{-4}\;{\rm g}\,{\rm cm}^{-3},\quad
\rho_{\rm vac}^{QCD}\sim 1.6\,10^{15}\;{\rm g}\,{\rm cm}^{-3},\quad
\rho_{\rm vac}^{Pl}\sim 2 \,10^{89}\;{\rm g}\,{\rm cm}^{-3}
\end{equation}
differs from the one measured from astronomical observations at cosmological scale
\begin{equation}\label{DensityL}
\rho_{\Lambda}=\frac{\Lambda}{8\pi{\rm G}}\sim h^{2}\,10^{-29}\;{\rm g}\,{\rm cm}^{-3}
\end{equation}
by $25$--$118$ orders of magnitude. Other estimations of this quantum effect from the viewpoint of standard Casimir energy
calculation scheme\cite{Zeldovich67} provide us with discrepancies of $\sim 37$ orders of magnitude\cite{Cherednikov02}.

A similar problem happens when 
\begin{equation}\label{LambdaV}
\Lambda_{\rm vac}=8\pi{\rm G}\rho_{\rm vac}
\end{equation}
is interpreted as a cosmological constant. Indeed, if the quantum field theory which provides us with an estimate of $\rho_{\rm vac}$ is
correct then the distance from which the gravity becomes repulsive in the sun neighborhood ranges from
$r_{\circ}^{EW}\sim2\,10^{-2}h^{-2/3}\,{\rm a.u.}$ down to $r_{\circ}^{Pl}\sim3\,10^{-11}h^{-2/3}\,{\rm \AA}$ depending on the quantum
field theory, see Eq.\,(\ref{r0}). Obviously, such results are not consistent with the observations.

Another version of the cosmological constant problem points out a fine tuning problem. It consists on arguing on the smallness of 
$\Lambda=\Lambda_{\rm vac}+\Lambda_{\circ}$, interpreted as an effective cosmological constant, where $\Lambda_{\circ}$ stands for a
bare cosmological constant in EinsteinÕs field equations.

\subsection{Understanding the acceleration of the cosmological expansion}

The observations show that the dynamics of the cosmological expansion agrees with the Friedmann-Lema\^{\i}tre-Gamov solution. It
describes an uniform distribution of pressureless matter and CMB radiation with a black-body spectra, the field equations are given by
Eq.\,(\ref{FundEq}) with $n\leq1$. The present values of related densities are $\rho_{\rm m}=3h^{2}\,10^{-30}\;{\rm g}\,{\rm cm}^{-3}$
(dark matter included) and $\rho_{\rm r}\sim 5h^{2}\,10^{-34}\;{\rm g}\,{\rm cm}^{-3}$. Their comparison to the expected vacuum energy
density $\rho_{\rm vac}$ shows that if quantum fluctuations intervene in the dynamics of the cosmological expansion then their
contribution prevails over the other sources (by $26$--$119$ orders of magnitude today). Such an hypothesis provides us with a vacuum
dominate cosmological expansion since primordial epochs. Therefore, one might ask whether such disagreements with observations
can be removed by taking into account higher order terms in Eq.\,(\ref{FundEq}). With this in mind, for describing the dynamics of
structures at scales where gravitational repulsion ($\Lambda> 0$) is observed, it is more convenient to use adapted units of time
$l_{g}$ and of mass $m_{g}$ defined as follows
\begin{equation}\label{GravitUnits}
l_{g}=1/\sqrt{\Lambda}\sim h^{-1}\,10^{28}\;{\rm cm},\qquad m_{g}=1/(8\pi G\sqrt{\Lambda})\sim 4h^{-1}\,10^{54}\;{\rm g}
\end{equation}
herein called {\em gravitational units\/}. They are defined such that the field equations read in a normalized
form
\begin{equation}\label{EqEinstein}
T_{\mu\nu}=-g_{\mu\nu}+S_{\mu\nu}+A_{2}F_{\mu\nu}^{(2)}+\ldots
\end{equation}
{\it i.e.\/} $A_{0}=A_{1}=1$, where the stress-energy tensor $T_{\mu\nu}$ accounts for the distribution of gravitational sources. It is important to
note that, with gravitational units, Planck constant reads
\begin{equation}\label{PlanckConstant}
\hbar\sim
10^{-120}
\end{equation}
Indeed, such a tiny value as {\em quantum action unit\/} compared to $\hbar=1$ when quantum units are used instead, shows clearly that
Eq.\,(\ref{EqEinstein}) truncated at $n\leq 1$ is not adapted for describing quantum physics\cite{Triay02,Triay04}. This is the main reason why it is
hopeless to give a quantum status to $\Lambda$\cite{SouriauTriay97}. As approximation, because of dimensional analysis described above, the
contribution of higher order invariants being the more significant as the density is large, Eq.\,(\ref{EqEinstein}) can be splited up with respect
to scale into two equations systems. The first one corresponds to terms of order $n<2$ (the usual Einstein equation with $\Lambda$) and the second
one
\begin{equation}\label{EqEinstein2}
T_{\mu\nu}^{\rm vac}= A_{2}F_{\mu\nu}^{(2)}+\ldots
\end{equation}
stands for the field equations describing the effect of quantum fluctuations on the gravitational field at an appropriated scale (quantum),
interpreted as correction of the RW metric $g_{\mu\nu}$. The identification of constants $A_{n}$ ({\it e.g.\/},
$A_{2}=\hbar$) and the derivation of tensors $F_{\mu\nu}^{(n)}$ with $n\geq2$ requires to model gravitational phenomena at quantum scale, see
{\it e.g.\/}\cite{ThooftVeltmann74,Stelle77}. Unfortunately, the state of the art does not allow yet to provide us with a definite answer for
defining the right hand term of Eq.~(\ref{EqEinstein2}), see {\it e.g.\/}\cite{Rovelli03}.

\section{Conclusion}
To rescale the field equations for describing the cosmological expansion prevents us to assume the vacuum acting as a cosmological constant.  As a
consequence, one understands that such an interpretation turns to be the origin of the cosmological constant problem\footnote{In other words, CCP is
the price to pay for identifying $\Lambda_{\rm vac}$ to the cosmological constant.}. Because the understanding of quantum gravity is still an ongoing
challenge, the correct field equations describing the contribution to gravity of quantum fluctuations are not yet established. However, the
dimensional analysis shows that the related gravitational effects are expected at small (quantum) scales and do not participate to the general
expansion of the universe according to observations.

\end{document}